\documentclass[
aps,prd,
12pt,
nopreprintnumbers,
showpacs,
eqsecnum,
nofootinbib
]{revtex4-1}

\usepackage{graphicx}

\begin{document}

\title{Critical Combinations of Higher Order Terms in Einstein--Maxwell
theory and Compactif{}ication}
\author{Nahomi Kan}\email[]{kan@gifu-nct.ac.jp}
\affiliation{
Gifu National College of Technology,
Motosu-shi, Gifu 501-0495, Japan
}
\author{Kiyoshi Shiraishi}\email[]{shiraish@yamaguchi-u.ac.jp}
\affiliation{
Graduate School of Science and Engineering, Yamaguchi University,
Yamaguchi-shi, Yamaguchi 753--8512, Japan}
\date{\today}

\begin{abstract}
We discuss the role of a particular combination of higher derivative terms
in higher dimensional theories, especially in the background of
spontaneous compactification. We find that the special extension of
Einstein--Maxwell theory with higher order terms admits interesting
cosmological solutions. 
\end{abstract}



\maketitle

\section{Introduction}
\label{sec1}

The gravitational theory with higher derivative terms
\cite{Stelle1,Stelle2} is one of the most interesting subjects to study in
physics for many years. The treatment of Einstein gravity with
perturbative field quantization naturally leads to inclusion of such
nonlinear terms in the spacetime curvatures in effective actions. 

The generalization of Einstein gravity has been developed extensively by
many authors.
The Lanczos--Lovelock gravity
\cite{Lanczos1,Lanczos2,Lovelock1,Lovelock2}%
\footnote{For a review, see \cite{PK}}, which is dictated by the
action with the dimensionally continued Euler forms, is defined in
a generic dimensional spacetime and it describes a massless spin-two
fluctuation despite of the existence of higher order terms in its action.
In a few decades ago, it is discovered that the effective field theory
arising from  string theory involves similar terms at least the
lowest-order correction to the Einstein-Hilbert term \cite{Zumino,Zwi}.

The other specific extension is found in the study on so-called critical
gravity \cite{LP,DLLP,LPP} (although the word ``critical''
originally implied the emergence of Log gravity \cite{AF,BHRT,JNZ}). In
such a theory of gravity, the behavior of spin-two modes is described by
the Lagrangian of special combination of higher order terms in the Ricci
tensor and the scalar curvature. There is a ghost field but no spin-zero
mode in the theory. In the three dimensional spacetime, the model
governed by a similar Lagrangian expresses the new massive gravity
\cite{NMG1,NMG2}.

In this paper, we consider incorporation of Maxwell fields to the
(next-to) critical gravity.
The critical gravity has been studied under the assumption of the
maximally symmetric background spacetime. The higher derivative theory on
the other  type of background fields is worth studying.
Inclusion of a non-vanishing flux field brings about the
partial compactification with a maximally symmetric extra space
and might provide an interesting scenario for a very early stage of the
universe.

The assumption of the partially symmetric solution can yield another
possibility. It is feasible to attach the other type of higher order terms
to the Lagrangian which realizes the spontaneous compactification.
Because of the symmetry, use of the Riemann tensor is possible in the
additional terms. They are expressed by special combinations of
Gauss-Bonnet terms and Horndeski's generalized Maxwell term
\cite{Horndeski1}. This can induce a second order field equation
displaying the cosmological evolution of the background geometry.%
\footnote{The applications of the Horndeski's vector-tensor theory have
been discussed in four dimensional \cite{BTY,JDHT} and also in
six dimensional cosmology \cite{YS}.}

This paper is organized as follows.
In Sec.~\ref{sec2}, we build the Lagrangian of our first model,
which involves the quadratic terms of the Ricci tensor, the Ricci scalar
and the two-form flux field, in a similar way revealed in the study of
critical gravity. The spontaneous compactification in the first model is
studied in Sec.~\ref{sec3} and cosmological solutions are examined in
Sec.~\ref{sec4}. In Sec.~\ref{sec5}, we present the second model
containing the quadratic term of the Riemann tensor. We also investigate
the cosmological solution qualitatively in this model. The last section is
devoted to summary and prospects.

Throughout the present paper, we deal with our models at the classical
level.

\section{a critical modification of Einstein-Maxwell theory}
\label{sec2}
In the work on critical gravity, it is found that the Lagrangian of the
specific combination of curvatures yields a spin-two massless graviton
and a ghost mode; there appears no scalar mode as in Einstein gravity.
In this section, we propose a model of higher order extension of
Maxwell--Einstein theory. We will follow the method for constructing the
Lagrangian examined in our previous work
\cite{KKS1,KKS2}. 

First, we start with the action for the Einstein
gravity and electromagnetism in $D$ dimensions:
\begin{equation}
I_0=\int d^Dx\sqrt{-g}L_0=\int
d^Dx\,\sqrt{-g}\,\left[R-F_{MN}F^{MN}-\Lambda\right]
\,,
\end{equation}
where $R$ and $\Lambda$ denote the scalar curvature obtained from
the metric $g_{MN}$ and the cosmological constant, respectively. The field
strength
$F_{MN}$ is defined by
$F_{MN}\equiv\partial_M A_N-\partial_N A_M$ with a vector field $A_M$
and $F^2\equiv F_{MN}F^{MN}$.
The suffices $M, N, \dots$ vary over $0, 1,\dots, D-1$.

Applying the variational principle to the action
results in the following equations of motion:
\begin{equation}
T_{MN}=0\,
\end{equation}
with
\begin{equation}
T_{MN}\equiv\frac{1}{\sqrt{-g}}\frac{\delta I_0}{\delta
g^{MN}}=R_{MN}-\frac{1}{2}Rg_{MN}
-2\left(F^2_{MN}-\frac{1}{4}F^2g_{MN}\right)+\frac{1}{2}\Lambda
g_{MN}\,,
\end{equation}
where $F^2_{MN}\equiv F_{MP}F_N{}^P$,
and
\begin{equation}
J^N\equiv\nabla_M F^{MN}=0\,.
\end{equation}

Next, we construct a Lagrangian of our model as follows:
\begin{equation}
L=\alpha L_0-\beta\, T^{MN}\Delta_{MNPQ}T^{PQ}
\,,
\end{equation}
where
\begin{equation}
\Delta_{MNPQ}\equiv\frac{1}{2}
(g_{MP}g_{NQ}+g_{MQ}g_{NP})-\frac{1}{D-1}
g_{MN}g_{PQ}\,.
\end{equation}
Note that the solution of $T_{MN}=0$ is the solution for the
equation of motion derived from $L$, provided that the field
equation for the vector field is satisfied (though the stability is not
necessarily guaranteed).%
\footnote{Generally speaking, we can add the term $\propto J_MJ^M$
to our Lagrangian, but we restrict ourselves to the case without such a
term in the present paper.}

The Lagrangian $L$ can then be written as
\begin{eqnarray}
L&=&\alpha\left[R-F^2-\Lambda\right] \nonumber \\
&-&\beta\left[R_{MN}R^{MN}-\frac{D}{4(D-1)}R^2-4F^2_{MN}R^{MN}
+\frac{D+2}{2(D-1)}F^2R\right.\nonumber \\
& &\quad+4F^2_{MN}F^2{}^{MN}-\frac{D+8}{4(D-1)}(F^2)^2\nonumber \\
& &\quad\left.+\frac{\Lambda}{2(D-1)}\left\{(D-2)R-(D-4)F^2-\frac{D}{2}
\Lambda\right\}\right]
\,.
\label{ours}
\end{eqnarray}

Incidentally, we can pick up some `critical' cases:
If $\alpha=\frac{D}{4(D-1)}\beta\Lambda$, the cosmological constant is
zero in the $D$ dimensional spacetime; If
$\alpha=\frac{D-2}{2(D-1)}\beta\Lambda$, the Einstein--Hilbert term $(
R)$ is absent from the $D$ dimensional action;
If
$\alpha=\frac{D-4}{2(D-1)}\beta\Lambda$, the pure Maxwell term $(F^2)$
is absent from the $D$ dimensional action.

We should note that the auxiliary field method
\cite{NMG1,NMG2,BHRT,KKS2}, which can be used in
critical higher order gravity, also leads to the additional terms, such as
\begin{equation}
L'=\alpha L_0-\beta(2T_{MN}S^{MN}-S_{MN}S^{MN}+S^2)\,,
\end{equation}
where $S^{MN}$ is an auxiliary symmetric tensor field and $S\equiv
S^{MN}g_{MN}$.
Varying the auxiliary field $S^{MN}$, we obtain
\begin{equation}
T_{MN}=S_{MN}-Sg_{MN}\,,
\label{ae}
\end{equation}
or, solving that with respect to $S_{MN}$, we get
\begin{equation}
S_{MN}=T_{MN}-\frac{1}{D-1}Tg_{MN}=\Delta_{MNPQ}T^{PQ}\,.
\end{equation}
Therefore, one can see the equivalence between the Lagrangians $L$ and
$L'$. We can derive the linearized field equation using the action
$I'=\int d^D x\, L'$.

Now, we investigate the behavior of linear fluctuations around the
background geometry, which satisfies $T^{MN}=0$.
We decompose the metric as follows:
\begin{equation}
g_{MN}=\bar{g}_{MN}+h_{MN}\,.
\end{equation}
The indices are raised and lowered by the background metric
$\bar{g}_{MN}$. Then the trace of the fluctuation is expressed as
\begin{equation}
h\equiv\bar{g}^{MN}h_{MN}\,.
\end{equation}
For conveniences, we write down the curvature tensors up to the
linear order in $h_{MN}$ here \cite{BC}:
\begin{eqnarray}
R^{MN}{}_{PQ}&=&\bar{R}^{MN}{}_{PQ}-\frac{1}{2}\bar{R}^{ML}{}_{PQ}h_L^N
-\frac{1}{2}\bar{R}^{NL}{}_{PQ}h_L^M\nonumber \\
& &+\frac{1}{2}(-\bar{\nabla}_P\bar{\nabla}^M h^N_Q+
\bar{\nabla}_P\bar{\nabla}^N h^M_Q
+\bar{\nabla}_Q\bar{\nabla}^M h^N_P-
\bar{\nabla}_Q\bar{\nabla}^N h^M_P)\,, \\
R^M_N&=&\bar{R}^M_N-\bar{R}^{MP}{}_{NQ}h_P^Q+\frac{1}{2}\bar{R}^M_Lh^L_N
-\frac{1}{2}\bar{R}^L_Nh_L^M\nonumber \\
& &+\frac{1}{2}(-\bar{\nabla}^2 h^M_N-
\bar{\nabla}_N\bar{\nabla}^M h+
\bar{\nabla}^M\bar{\nabla}_L h^L_N
+\bar{\nabla}_N\bar{\nabla}_L h^{LM})\,,
\label{ricci}
\\
R&=&\bar{R}-\bar{R}^{MN} h_{MN}
+\bar{\nabla}_M\bar{\nabla}_N h^{MN}-\bar{\nabla}^2
h\,,
\label{scalar}
\end{eqnarray}
where the nabla stands for the covariant derivative and the barred symbols
show that they are constructed from the background metric $\bar{g}_{MN}$.

Assuming the background fields satisfies
$\bar{T}_{MN}=\bar{R}_{MN}-\frac{1}{2}\bar{R}\bar{g}_{MN}
-2\left(\bar{F}^2_{MN}-\frac{1}{4}\bar{F}^2\bar{g}_{MN}\right)+\frac{1}{2}\Lambda
\bar{g}_{MN}=0$, we can use
\begin{eqnarray}
\delta
& &(2S^{PQ}T_{PQ}-S_{PQ}S^{PQ}+S^2)=2S^{PQ}\delta T_{PQ}\nonumber \\
& &\qquad\qquad+(\mbox{higher order
terms in small fluctuations from the background})\,,
\end{eqnarray}
to obtain the field equation at the linearized order:
\begin{eqnarray}
\frac{1}{\sqrt{-g}}\frac{\delta I'}{\delta g_{MN}}=0&\rightarrow&
\nonumber \\
 -\alpha
T^{MN}&-&\beta\left[-\bar{\nabla}^2S^{MN}-\bar{\nabla}^M\bar{\nabla}^NS
+\bar{\nabla}^M\bar{\nabla}_LS^{LN}+\bar{\nabla}^N\bar{\nabla}_LS^{LM}\right.
\nonumber \\
&
&-\bar{\nabla}_P\bar{\nabla}_QS^{PQ}\bar{g}^{MN}+\bar{\nabla}^2S\bar{g}^{MN}
-2\bar{R}_P{}^M{}_Q{}^NS^{PQ}\nonumber \\
& &
+\bar{R}_{PQ}S^{PQ}\bar{g}^{MN}-\bar{R}S^{MN}+\bar{R}^M_LS^{LN}
+\bar{R}^N_LS^{LM}+\Lambda S^{MN}\nonumber \\
&
&\left.+4\bar{F}^M{}_P\bar{F}^N{}_QS^{PQ}+\bar{F}^2S^{MN}
-2\bar{F}^2_{PQ}S^{PQ}\bar{g}^{MN}\right]=0\,,
\label{dg}
\end{eqnarray}
noting that $S^{MN}$ is in the linear order ($\bar{S}^{MN}=0$), and we
have used the condition $\bar{T}_{MN}=0$ for the background field.

The generalized Maxwell equation in the linear order in
fluctuations takes the form
\begin{equation}
\frac{1}{\sqrt{-g}}\frac{\delta I'}{\delta A_{N}}=0\rightarrow 2\alpha
\bar{\nabla}_Mf^{MN}
-2\beta\bar{\nabla}_M\left[2\bar{F}^M{}_PS^{PN}-2\bar{F}^N{}_PS^{PM}
-\bar{F}^{MN}S\right]=0\,,
\end{equation}
where $f_{MN}=F_{MN}-\bar{F}_{MN}$.

The linearized equation obtained from (\ref{ae}) is
\begin{eqnarray}
T^{MN}&=&S^{MN}-g^{MN}S\rightarrow \nonumber \\
& &
\frac{1}{2}\left[-\bar{\nabla}^2h^{MN}-\bar{\nabla}^M\bar{\nabla}^Nh
+\bar{\nabla}^M\bar{\nabla}_Lh^{LN}+\bar{\nabla}^N\bar{\nabla}_Lh^{LM}\right.
\nonumber \\
&
&-\bar{\nabla}_P\bar{\nabla}_Qh^{PQ}\bar{g}^{MN}+\bar{\nabla}^2h\bar{g}^{MN}
-2\bar{R}_P{}^M{}_Q{}^Nh^{PQ}\nonumber \\
& &
\left.+\bar{R}^{PQ}h_{PQ}\bar{g}^{MN}-\bar{R}h^{MN}+\bar{R}^M_Lh^{LN}
+\bar{R}^N_Lh^{LM}+\Lambda h^{MN}\right]\nonumber \\
&
&+2\bar{F}^M{}_P\bar{F}^N{}_Qh^{PQ}+\frac{1}{2}\bar{F}^2h^{MN}
-\bar{F}^2_{PQ}h^{PQ}\bar{g}^{MN}\nonumber \\
&
&-2\bar{F}^{PM}f_{P}{}^N-2\bar{F}^{PN}f_{P}{}^M
+\bar{F}^{PQ}f_{PQ}\bar{g}^{MN}=S^{MN}-\bar{g}^{MN}S\,.
\end{eqnarray}

The important observation is that, using (\ref{ae}),
the field equation (\ref{dg}) becomes
\begin{eqnarray}
& &-\bar{\nabla}^2S^{MN}-\bar{\nabla}^M\bar{\nabla}^NS
+\bar{\nabla}^M\bar{\nabla}^LS^{LN}+\bar{\nabla}^N\bar{\nabla}^LS^{LM}
-\bar{\nabla}_P\bar{\nabla}_QS^{PQ}\bar{g}^{MN}
+\bar{\nabla}^2S\bar{g}^{MN}\nonumber \\
& &
+\frac{\alpha}{\beta}
(S^{MN}-\bar{g}^{MN}S)-2\bar{R}_P{}^M{}_Q{}^NS^{PQ}
+\bar{R}_{PQ}S^{PQ}\bar{g}^{MN}-\bar{R}S^{MN}+\bar{R}^M_LS^{LN}
+\bar{R}^N_LS^{LM}+\Lambda S^{MN}\nonumber \\
&
&+4\bar{F}^M{}_P\bar{F}^N{}_QS^{PQ}+\bar{F}^2S^{MN}
-2\bar{F}^2_{PQ}S^{PQ}\bar{g}^{MN}=0\,.
\label{fp}
\end{eqnarray}
In the $D$ dimensional Minkowski vacuum, this equation is just the
Fierz-Pauli equation \cite{FP} for the spin-two wave with mass-squared
$m^2=\frac{\alpha}{\beta}$.
Therefore we recognize that $S^{MN}$ corresponds to the massive ghost
field. The postulate of no tachyonic ghost requires
$\frac{\alpha}{\beta}>0$ in the present model.

The linear fluctuation from the background fields can be argued by using
above equations. 
In the next section, we consider the solution for spontaneous
compactification of spacetime as the background.

\section{Spontaneous compactification in the critically modified higher
order Einstein-Maxwell theory}
\label{sec3}
In this section, we study the partial compactification of space with a
non-trivial flux in our model.
The extra dimensions are considered to be compactified into a sufficiently
small size. A maximally symmetric solution in our model is
compactification of the form
$M_{D-2}\otimes S^2$, where $M_{D-2}$ is the $(D-2)$ dimensional Minkowski
spacetime and
$S^2$ is a two dimensional sphere, as in the work of Randjbar-Daemi, Salam
and Strathdee (RSS) \cite{RSS} where $D=6$ is considered.  We consider
the solution for flat $D-2$ dimensional spacetime and study its
classical stability in this section, though other maximally symmetric
spacetimes are also interesting. Our notation is that the suffices $\mu,
\nu,\dots$ run over $0, 1, 2, D-3$ and the suffices $m, n, \dots$ are used
for the extra dimensions $D-2$ and $D-1$.

As stated in the previous section, the solution of $T_{MN}=0$ and $J_M=0$
is a solution in our model.
A suitable ansatz for the flux field is 
\begin{equation}
F=\frac{1}{2}F_{MN}dx^M\wedge dx^N=\frac{1}{2}q\varepsilon_{mn}
dx^m\wedge dx^n\,,
\end{equation}
where $\varepsilon_{mn}$ is the antisymmetric symbol and the constant $q$
is the strength of the two-form flux.%
\footnote{It can be checked that this
configuration is the solution of our model.}
 This monopole flux yields the relation
\begin{equation}
F^2_{mn}=\frac{1}{2}F^2g_{mn}\,.
\end{equation}

Then, the equation $T_{MN}=0$ can be decomposed to
\begin{eqnarray}
R_{\mu\nu}&=&-\frac{1}{D-2}F^2g_{\mu\nu}+\frac{1}{D-2}\Lambda
g_{\mu\nu}\,,\label{eom1}\\
R_{mn}&=&2F^2_{mn}-\frac{1}{D-2}F^2g_{mn}+\frac{1}{D-2}\Lambda g_{mn}
=\frac{D-3}{D-2}F^2g_{mn}+\frac{1}{D-2}\Lambda g_{mn}\,,
\label{eom2}
\end{eqnarray}
for the spacetime of a direct product.
As for the flat $D-2$ dimensional spacetime, where $R_{\mu\nu}=0$,
we must take a fine-tuning of parameters as 
\begin{equation}
\bar{F}^2=\Lambda>0\,,
\end{equation}
and therefore, the solution for the background fields reads
\begin{equation}
\bar{R}=2\bar{F}^2\,,\quad \bar{R}_{mn}=\Lambda
\bar{g}_{mn}=2\bar{F}^2_{mn}\,,\quad
\bar{R}^{mn}{}_{pq}=\Lambda(\delta^m_p\delta^n_q-\delta^p_q\delta^n_p)=
2\bar{F}^{mn}\bar{F}_{pq}\,.
\label{bs1}
\end{equation}

Substituting this compactified background solution into Eq.~(\ref{fp}),
we obtain
\begin{eqnarray}
& &-\bar{\nabla}^2S^{MN}-\bar{\nabla}^M\bar{\nabla}^NS
+\bar{\nabla}^M\bar{\nabla}^LS^{LN}+\bar{\nabla}^N\bar{\nabla}^LS^{LM}
-\bar{\nabla}_P\bar{\nabla}_QS^{PQ}\bar{g}^{MN}
+\bar{\nabla}^2S\bar{g}^{MN}\nonumber \\
& &+\bar{R}^M_LS^{LN}+\bar{R}^N_LS^{LM}+\frac{\alpha}{\beta}
(S^{MN}-\bar{g}^{MN}S)=0\,.
\label{ss}
\end{eqnarray}

To show complete particle spectrum for the ghost field is rather
complicated, but we can see that the lowest mass-squared of the ghost is
$\alpha/\beta$ for symmetric tensor fields (which do not couples to the
Ricci curvature of the extra space) and vector bosons of the Kaluza-Klein
origin (which come from the zero modes of the Lichnerowicz operator on
$S^2$,
$-\bar{\nabla}^2+\bar{R}_{mn}$). It is also remarkable that the other
graviton and vector field fluctuations is absent in the linearized
equation (\ref{ss}). 

Thus, taking the analysis by RSS for other fluctuation modes \cite{RSS}
into account, in the classical and linearized analysis, we find that the
RSS background solution $M_{D-2}\otimes S^2$ is stable if
$\alpha/\beta>0$, because there is no growing mode.%
\footnote{Of course, there remain the problem of nonlinear instability
and the ghost problem in quantum physics.}

Before closing this section, we note that the effective $D-2$ dimensional
Newton constant, which can be found by the coefficient of the Ricci
scalar of the $D-2$ dimensional spacetime in the action, is not affected
by the higher order terms in our model and is independent of $\beta$.   
It might be related to the absence of the graviton and the ghost modes
at the linearized level mentioned above.

\section{de Sitter solutions and cosmology in the model}
\label{sec4}
In this section, we perform the analysis on the
stability of the compactification and the cosmological
solution in our model.
We consider an effective potential $V$ for the radius
of extra space $S^2$  \cite{Wetterich1} in order to study them.

We can express the Riemann tensor of the $S^2$ as
\begin{equation}
R^{mn}{}_{pq}=\frac{1}{b^2}(\delta^m_p\delta^n_q-\delta^m_q\delta^n_p)\,,
\end{equation}
where $b$ denotes the radius of the compactified sphere $S^2$.
Then, the equations (\ref{eom1},\ref{eom2}) with $R_{\mu\nu}=0$
impose \cite{RSS}
\begin{equation}
q^2=\frac{1}{2\Lambda}\,,\quad b^2=\frac{1}{\Lambda}\,,
\end{equation}
where the first equation is the fine-tuning condition between the
parameters.

The effective potential $V$ as a function of $b$,
the scale of the extra space,
 is obtained from
replacing the background fields by functions of $b$ in $-I$ with
\begin{equation}
R^m_n=\frac{1}{b^2}\delta^m_n\,,\quad
(F^2){}^m_n=\frac{q^2}{b^4}\delta^m_n=\frac{1}{2\Lambda b^4}\delta^m_n\,,
\end{equation}
and $\sqrt{\det \bar{g}_{mn}}\propto b^2$.
The effective potential for the RSS model is calculated from $-I_0$
and is given as
\begin{equation}
V_0(\Lambda b^2)=-b^2\left(\frac{2}{b^2}-\frac{2q^2}{b^4}-\Lambda\right)=
-2+\frac{1}{\Lambda b^2}+\Lambda b^2\,,
\end{equation}
and then, the effective potential for our model of which Lagrangian is
described by (\ref{ours}) is
\begin{eqnarray}
V(y)&=&\alpha\left[-2+\frac{1}{y}+y\right]\nonumber 
\\
& &+\frac{\beta\Lambda}{4(D-1)}
\left[4(D-2)-D\,y+\frac{2D}{y}-\frac{12(D-2)}{y^2}+\frac{7D-16}{y^3}\right]
\nonumber \\
&=&\frac{(y-1)^2}{4(D-1)y^3}[(7D-16)\beta+2(D-4)\beta
y+(4(D-1)\alpha-D\beta)y^2]\,,
\end{eqnarray}
with $y\equiv\Lambda b^2$.
A minimum of the potential is found at $y=1$ and attains $V(1)=0$.
The stability of the radius of the extra sphere is determined by the sign
of the second derivative of the potential:
\begin{equation}
V''(1)=
2\left(\alpha+2\frac{D-3}{D-1}\beta\Lambda\right)\,.
\label{v2}
\end{equation}
If $D>3$ and $\alpha, \beta>0$, the scale $b=1/\sqrt{\Lambda}$ is
found to be stable. We also find that it is impossible to create another
potential minimum satisfying $V=0$ by any tuning of $\beta/\alpha$
in the model.

At a large value of $b$, $V$ can be indefinitely negative provided that
$\frac{D}{4(D-1)}\beta\Lambda-\alpha>0$. In this case, the $D$ dimensional
cosmological constant becomes negative. This may not indicate instability
of compactification against a large fluctuation, because the higher
derivative terms cause the coupling between the compactification scale
and the first and second derivatives of an expanding scale factor of the
rest of space. 

Therefore, we investigate the cosmological equation including the
dynamical scale factor and the compactification scale.
The metric is usually assumed as
\begin{equation}
ds^2=-dt^2+a^2(t)d\Omega_{D-3}^2+b^2(t)d\Omega_2^2\,,
\end{equation}
where $b(t)$ denotes the radius of the compactified sphere $S^2$
and $d\Omega_2^2$ is the line element an the unit sphere.
For the maximally symmetric $(D-3)$ dimensional space,
the line element is denoted $d\Omega_{D-3}^2$ here,
and its Riemann tensor is normalized as
$\tilde{R}^{ij}{}_{kl}=
k(\delta^i_k\delta^j_l-\delta^i_l\delta^j_k)$
with $k=0, \pm 1$, where $i, j,\dots=1,
2,\dots, D-3$. $a(t)$ is the scale factor of $D-3$ dimensional
homogeneous and isotropic space.

The Riemann curvature is computed with the metric as
\begin{eqnarray}
& &R^{0i}{}_{0j}=\frac{\ddot{a}}{a}\delta^i_j\,,\quad
R^{0m}{}_{0n}=\frac{\ddot{b}}{b}\delta^m_n\,,\quad
R^{im}{}_{jn}=\frac{\dot{a}}{a}\frac{\dot{b}}{b}
\delta^i_j\delta^m_n\,,\nonumber \\
& &R^{ij}{}_{kl}=\left(\frac{\dot{a}^2+k}{a^2}\right)
\left(\delta^i_k\delta^j_l-\delta^i_l\delta^j_k\right)\,,\quad
R^{mn}{}_{pq}=\left(\frac{\dot{b}^2+1}{b^2}\right)
\left(\delta^m_p\delta^n_q-\delta^m_q\delta^n_p\right)\,,
\end{eqnarray}
where the dot $(\dot{~})$ denotes the derivative with respect to the
cosmic time $t$.

Now we can obtain the effective action of the scale factor $a(t)$ and the
compactification scale $b(t)$ for the RSS model as
\begin{eqnarray}
& &I_0\rightarrow\nonumber \\
& & \int dt\, a^{D-3}b^2\left[
2(D-3)\frac{\ddot{a}}{a}+4\frac{\ddot{b}}{b}
+4(D-3)\frac{\dot{a}}{a}\frac{\dot{b}}{b}
+(D-3)(D-4)\frac{\dot{a}^2+k}{a^2}\right.\nonumber
\\
& &\left.\qquad\qquad\qquad+2\frac{\dot{b}^2+1}{b^2} -\frac{1}{\Lambda
b^4}-\Lambda
\right]\nonumber \\
&\sim&\int dt\, a^{D-3}b^2\left[
-4(D-3)\frac{\dot{a}}{a}\frac{\dot{b}}{b}
+(D-3)(D-4)\frac{-\dot{a}^2+k}{a^2}\right.\nonumber
\\
& &\left.\qquad\qquad\qquad+2\frac{-\dot{b}^2+1}{b^2} -\frac{1}{\Lambda
b^4}-\Lambda
\right]\equiv \tilde{I}_0\,,
\end{eqnarray}
where the over-all constant which comes from
the volume of space has been omitted. The action has been integrated by
parts after the symbol `$\sim$'.

The effective action for our model can be written, by adopting the
auxiliary fields, as
\begin{eqnarray}
& &I'\rightarrow \tilde{I}_0\nonumber \\
& &-\beta\int dt\, a^{D-3}b^2\left\{2s_0\left[-\frac{(D-3)(D-4)}{2}
\frac{\dot{a}^2+k}{a^2}
-2(D-3)\frac{\dot{a}}{a}\frac{\dot{b}}{b}
-\frac{\dot{b}^2+1}{b^2}+\frac{1}{2\Lambda
b^4}+\frac{1}{2}\Lambda\right]\right.\nonumber \\
&
&+2(D-3)s_a\left[-(D-4)\frac{\ddot{a}}{a}-2\frac{\ddot{b}}{b}-2(D-4)
\frac{\dot{a}}{a}\frac{\dot{b}}{b}-\frac{(D-4)(D-5)}{2}
\frac{\dot{a}^2+k}{a^2}-\frac{\dot{b}^2+1}{b^2}+\frac{1}{2\Lambda
b^4}+\frac{1}{2}\Lambda\right]\nonumber
\\ & &+4s_b\left[-(D-3)\frac{\ddot{a}}{a}-\frac{\ddot{b}}{b}-(D-3)
\frac{\dot{a}}{a}\frac{\dot{b}}{b}-\frac{(D-3)(D-4)}{2}
\frac{\dot{a}^2+k}{a^2}-\frac{1}{2\Lambda
b^4}+\frac{1}{2}\Lambda\right]\nonumber \\
&
&\left.-\left[s_0^2+(D-3)s_a^2+2s_b^2\right]+\left[
s_0+(D-3)s_a+2s_b\right]^2\right\}\,,
\end{eqnarray}
where we set $S^0_0=s_0$, $S^i_j=s_a\delta^i_j$ and
$S^m_n=s_b\delta^m_n$.
Here, the equations of motion are omitted, but
they can be obtained from the above action.%
\footnote{The Euler equations can be obtained from the action using the
command {\tt EulerEquations} in the add-on {\tt
Calculus`VariationalMethods`} for {\tt Mathematica}\textregistered
\cite{Mathematica}.  }

We seek for the de Sitter ($\otimes S^2$) solution,
where
$k=0$, $a(t)=e^{H_0t}$ with constant $H_0$, and $b=b_0=$constant, are
taken as an ansatz. In this case,  all the equations of motion become
algebraic equations. With the ansatz, the auxiliary variables are
expressed as
\begin{eqnarray}
s_0=s_a&=&\frac{3-2\Lambda b_0^2+[D(D-3)H_0^2-\Lambda]\Lambda
b_0^4}{2(D-1)\Lambda b_0^4}\,,\nonumber \\
s_b&=&\frac{5-2D+2(D-2)\Lambda b_0^2-[(D-2)(D-3)H_0^2+\Lambda]\Lambda
b_0^4}{2(D-1)\Lambda b_0^4}\,.
\end{eqnarray}
Substituting these auxiliary relations to the other equations, 
we obtain the following two equations at last:
\begin{eqnarray}
& &-(7D-16)\beta+12(D-2)\beta\Lambda b_0^2\nonumber \\
&
&-2\left\{2(D-1)\alpha+[(D+2)(D-3)(D-4)H_0^2+D\Lambda]\beta\right\}
\Lambda b_0^4\nonumber \\
& &+4\left\{2(D-1)\alpha+[D(D-3)(D-4)H_0^2-(D-2)\Lambda]
\beta\right\}\Lambda^2
b_0^6\nonumber \\
&
&+\left\{(D-3)^2(D-6)(D^2-6D+4)H_0^4\beta+2(D-3)(D-4)H_0^2
[2(D-1)\alpha-(D-2)\beta\Lambda]\right.\nonumber \\
& &
\quad\left.+[-4(D-1)\alpha+D\beta\Lambda]
\Lambda\right\}\Lambda^2
b_0^8=0\,,
\label{se1}
\end{eqnarray}
and
\begin{eqnarray}
& &3(7D-16)\beta-24(D-2)\beta\Lambda b_0^2\nonumber \\
&
&+2\left\{2(D-1)\alpha+[(D+2)(D-2)(D-3)H_0^2+D\Lambda]\beta\right\}
\Lambda b_0^4\nonumber \\
&
&+\left\{(D-3)^2(D-2)(D^2-6D+4)H_0^4\beta+2(D-2)(D-3)H_0^2
[2(D-1)\alpha-(D-2)\beta\Lambda]\right.\nonumber \\
& &
\quad\left.+[-4(D-1)\alpha+D\beta\Lambda]
\Lambda\right\}\Lambda^2
b_0^8=0\,.
\label{se2}
\end{eqnarray}

The two sets of the solution for $b_0$ and $H_0$ can be found;
one is
\begin{equation} 
\mbox{(i)}\qquad H_0^2=0 \quad\mbox{and}\quad b_0^2=\frac{1}{\Lambda}\,,
\label{s1}
\end{equation}
 and another is
\begin{equation} 
\mbox{(ii)}\qquad H_0^2=\frac{D-4}{(D-3)^3}\Lambda \quad\mbox{and}\quad
b_0^2=\frac{D-3}{\Lambda}\,.
\label{s2}
\end{equation}
These values for solutions are the same as those of the RSS model
\cite{Okada1}, as expected (if $D=6$). In both cases, the values for
$s_0$,
$s_a$ and
$s_b$ are equal to zero.

Because the simultaneous equations (\ref{se1},\ref{se2}) are the fourth
order algebraic equations, there are other two sets of solutions%
\footnote{In general, the values for $s_0$, $s_a$ and
$s_b$ are not zero for such solutions.} 
at most.
One can find, for sufficiently small value for $\beta$, only two original
solutions (\ref{s1},\ref{s2}) are real solutions. For $D=6$, only two real
solutions (\ref{s1},\ref{s2}) exist for $\beta\Lambda/\alpha<\approx 2.8$.
As a model of the universe, the multiple de Sitter phases would have some
significance in cosmology.

We now study the behavior of the small fluctuation around the above exact
solutions.
In the neighborhood of (i), we set
\begin{equation}
b(t)=\frac{1}{\sqrt{\Lambda}}+\Delta b(t)\,,\quad
H(t)=\frac{\dot{a}}{a}=\Delta H(t)\,,
\end{equation} 
and $\Delta s_a=s_a$, and so on. Then, the linearized equations of motion
read
\begin{eqnarray}
& &(D-3)\Delta s_a+2\Delta s_b=0\,, \\
&
&(D-1)\Delta\ddot{s}_a+\left[2(D-3)\Lambda+(D-1)\frac{\alpha}{\beta}\right]
\Delta s_a=0\,, \label{res1}\\
& &2(D-1)\alpha\left[(D-2)\Delta\ddot{b}+2(D-4)\Lambda\Delta b\right]
\nonumber \\
&
&\quad+\frac{D-3}{\sqrt{\Lambda}}\left[(D-1)(D-2)\alpha+4\beta\Lambda
\right]\Delta s_a=0\,,\label{res2} \\
& &(D-1)\alpha\left[(D-2)\Delta\dot{H}-4\Lambda^{3/2}\Delta
b\right]\nonumber \\
& &\quad-\left[(D-1)(D-2)\alpha-2(D-3)\beta\Lambda\right]\Delta
s_a=0\,,
\\ & &(D-1) \Delta s_0= -2(D-3)
\frac{\beta\Lambda}{\alpha}
\Delta s_a\,.
\end{eqnarray} 
Note that for $\beta=0$, the equations give $\Delta s_a=\Delta s_b=\Delta
s_0=0$, $(D-2)\Delta\ddot{b}+2(D-4)\Lambda\Delta b=0$ and $(D-2)\Delta\dot{H}-4\Lambda^{3/2}\Delta
b=0$, which are equivalent to the equations for the cosmology of the RSS
model
\cite{Okada1}.

The oscillatory behavior of $\Delta b$ as well as $\Delta s_a$ is enough
interesting.
The variables $\Delta s_0$, $\Delta s_a$ and $\Delta s_b$ oscillate with
a different frequency from that of $\Delta b$, as seen from
(\ref{res1},\ref{res2}).%
\footnote{Note that the angular frequency of the oscillation of $\Delta
s_{0,a,b}$ is the square root of $\frac{V''(1)}{2\beta}$ in (\ref{v2}).}
Thus, the tuning of initial conditions realizes enough time for
accelerating expansion and a sufficient inflation can be achieved,
though we cannot tell the choice of initial state in the present
classical analysis of the model. 

Since the analysis of the perturbation from (ii) can be considered and
results in finding exponentially growing modes as found in \cite{Okada1},
we omit the indication of equations here. Some fine tuning gives an enough
expansion time for a sufficient inflation as the de Sitter phase also in
this case.

\section{spontaneous compactification in
Lovelock-Horndeski-nonlinear Maxwell theory}
\label{sec5}
In the solution for the spontaneous compactification in previous sections,
one has found the following relation in the background fields:
\begin{equation}
\bar{R}_{MNPQ}=2\bar{F}_{MN}\bar{F}_{PQ}\,.
\label{RFF}
\end{equation}
Similarly to the consideration in the previous sections,
we come to an idea that the Minkowski compactification of the RSS model
is attained in a higher derivative theory if the relation (\ref{RFF})
is reflected in the higher order correction term.

We then propose an additional higher order term
\begin{eqnarray}
L_2&=&\frac{1}{4}\delta^P_M{}^Q_N{}^R_K{}^S_L(R^{MN}_{PQ}-2F^{MN}F_{PQ})
(R^{KL}_{RS}-2F^{KL}F_{RS})\nonumber \\
&=&R^{MNPQ}R_{MNPQ}-4R^{MN}R_{MN}+R^2-4\left(
R^{MNPQ}F_{MN}F_{PQ}-4R^{MN}F^2_{MN}+RF^2\right)\nonumber \\
& &+4F^2{}^{MN}F^2_{MN}-8(F^2)^2\,,
\end{eqnarray}
where the generalized Kronecker's delta is defined by
\begin{equation}
\delta^{M_1 M_2\cdots M_p}_{N_1N_2\cdots N_p}
\equiv\left|\begin{array}{cccc}
\delta^{M_1}_{N_1} & \delta^{M_1}_{N_2} & \cdots &
\delta^{M_1}_{N_p} \\
\delta^{M_2}_{N_1} & \delta^{M_2}_{N_2} & \cdots &
\delta^{M_2}_{N_p} \\
\vdots & \vdots & \ddots & \vdots \\
\delta^{M_p}_{N_1} & \delta^{M_p}_{N_2} & \cdots &
\delta^{M_p}_{N_p} 
\end{array}\right|\,.
\end{equation}
Now, we write the model~2 by the Lagrangian (therefore, we call the
previous model the model~1)
\begin{equation}
L=\alpha L_0+\gamma L_2\,.
\end{equation}
Note that $L_2$ consists of the Lanczos--Lovelock Lagrangian
(Gauss--Bonnet term)
\cite{Lanczos1,Lanczos2,Lovelock1,Lovelock2} and the Horndeski's
vector-tensor term \cite{Horndeski1} and a nonlinear Maxwell term.

It is easy to see that adding $L_2$ to the RSS Lagrangian $L_0$ yields no
modification of linearized equation of motion on the RSS background
fields after gauge fixing. Therefore, the Kaluza-Klein modes remain
unchanged from those in the RSS model.
It is worth noting that the effective $D-2$ dimensional
Newton constant in model~2 is not affected
by the value of $\gamma$. 

Now, we study the cosmological evolution of the classical background
in the model. Although the Kaluza-Klein cosmologies with the Gauss--Bonnet
term or the generalized Euler invariant have been studied 
in many works including \cite{Madore,MH1,Henriques,Shiraishi,Arik,DB},
the case with the RSS-type compactification with the Gauss--Bonnet term
has been studied in a few papers including \cite{Mignemi}.%
\footnote{The compactification
with the Horndeski Lagrangian has been considered by one of the present
authors
\cite{YS}.}

As previously, we characterize the cosmological background as:
\begin{equation}
ds^2=-dt^2+a^2(t)d\Omega_{D-3}^2+b^2(t)d\Omega_2^2\,,
\end{equation}
and
\begin{equation}
F=\frac{1}{2}q\varepsilon_{mn}dx^m\wedge dx^n=
\frac{1}{2}\frac{1}{\sqrt{2\Lambda}}\varepsilon_{mn}dx^m\wedge dx^n\,.
\end{equation}
Then, the Lagrangian $L_2$ reduces to
\begin{eqnarray}
& &L_2=4(D-3)\frac{\ddot{a}}{a}\left[4(D-4)\frac{\dot{a}}{a}
\frac{\dot{b}}{b}+
(D-4)(D-5)\frac{\dot{a}^2+k}{a^2}+2\frac{\dot{b}^2+1}{b^2}
-2\frac{1}{\Lambda b^4}
\right]\nonumber \\
& &\qquad+8\frac{\ddot{b}}{b}\left[2(D-3)\frac{\dot{a}}{a}
\frac{\dot{b}}{b}+
(D-3)(D-4)\frac{\dot{a}^2+k}{a^2}\right]\nonumber \\
& &
\qquad+(D-3)(D-4)(D-5)(D-6)\left(\frac{\dot{a}^2+k}{a^2}\right)^2
\nonumber \\
&
&\qquad+8(D-3)(D-4)(D-5)\frac{\dot{a}}{a}\frac{\dot{b}}{b}\frac{\dot{a}^2+k}{a^2}
+8(D-3)(D-4)\frac{\dot{a}^2}{a^2}\frac{\dot{b}^2}{b^2}
\nonumber \\ 
& &\qquad
+4(D-3)(D-4)\frac{\dot{a}^2+k}{a^2}\left(\frac{\dot{b}^2+1}{b^2}
-\frac{1}{\Lambda b^4}\right)\,,
\end{eqnarray}
and we obtain the following action for $a$ and $b$:
\begin{eqnarray}
\tilde{I}_2&=&\int dt\, a^{D-3}b^2
\left\{(D-3)(D-4)(D-5)(D-6)\left[-\frac{1}{3}\frac{\dot{a}^4}{a^4}-2\frac{k}{a^2}
\frac{\dot{a}^2}{a^2}+\frac{k^2}{a^4}\right]\right.\nonumber \\
& &
\qquad\qquad\qquad+8(D-3)(D-4)(D-5)\left[-\frac{1}{3}\frac{\dot{a}^3}{a^3}
\frac{\dot{b}}{b}-\frac{k}{a^2}\frac{\dot{a}}{a}
\frac{\dot{b}}{b}\right]
\nonumber \\& &\qquad\qquad\qquad
+4(D-3)(D-4)\left[-\frac{\dot{a}^2}{a^2}\frac{\dot{b}^2+1}{b^2}
+\frac{k}{a^2}\frac{-\dot{b}^2+1}{b^2}\right]\nonumber \\&
&\qquad\qquad\qquad -4(D-3)(D-4)\frac{-\dot{a}^2+k}{a^2}\frac{1}{\Lambda
b^4}-16(D-3)\left.\frac{\dot{a}}{a}\frac{\dot{b}}{b}\frac{1}{\Lambda
b^4}\right\}\,.
\end{eqnarray}
Again, we omit exposition of equations of motion here.

We first explore the de Sitter phase, or the maximally symmetric
product spacetime. To find this, we set $k=0$,
$a(t)=e^{H_0t}$ and
$b(t)=b_0$, where $H_0$ and $b_0$ are constants.
We then obtain the simultaneous algebraic equations as follows:
\begin{eqnarray}
& &-[\alpha+4 (D-3)(D-4)H_0^2\gamma]+2\left[
\alpha+2(D-3)(D-4) H_0^2\gamma\right]\Lambda b_0^2\nonumber \\ &
&\quad+
\left[\left((D-3)(D-4)H_0^2-\Lambda\right)\alpha+(D-3)(D-4)(D-5)(D-6)
H_0^4\gamma\right]\Lambda b_0^4=0\,,\label{t1} \\& &
\alpha+4(D-2)(D-3)H_0^2\gamma \nonumber \\ & &\quad+
\left[\left((D-2)(D-3)H_0^2-\Lambda\right)\alpha+(D-2)(D-3)(D-4)(D-5)
H_0^4\gamma\right]\Lambda b_0^4=0\label{t2}\,,
\end{eqnarray}
Of course, the solution (i) (\ref{s1}) is a trivial solution of above
equations, while (ii) (\ref{s2}) is not a solution for $\gamma\ne 0$.
One solution of the simultaneous equations (\ref{t1},\ref{t2}) with
$H_0\ne 0$ exists for all positive $\gamma/\alpha$.
The value of $b_0 (>\frac{1}{\sqrt{\Lambda}})$ increases with the value of
$\gamma/\alpha$.
The existence of the de Sitter phase with large values for $H_0$ and
$b_0$ is suitable for the possible inflationary phase.

Next, we draw the phase diagram of the compactification scale $b$.
We consider the case with $k=0$ and $D=6$. FIG.~1 shows the phase space
spanned by $b$ and $\dot{b}$ for $\gamma\Lambda/\alpha=0, 1, 2$.
The small arrows in the figures indicate the flow in the phase space.
One can see that $b=1/\sqrt{\Lambda}$ is an attractor and the area of
attracting region increases with the value of $\gamma/\alpha$.%
\footnote{If ordinary matter is added, the dissipative motion becomes
more apparent in an expanding era \cite{Maeda1}.} We conclude that an
appropriate choice of the initial condition, which is in the flow passing
through near by the branch point given by the de Sitter solution of
(\ref{t1},\ref{t2}), might cause a long expansion
time.

\begin{figure}[ht]
\centering
\includegraphics[height=5cm]
{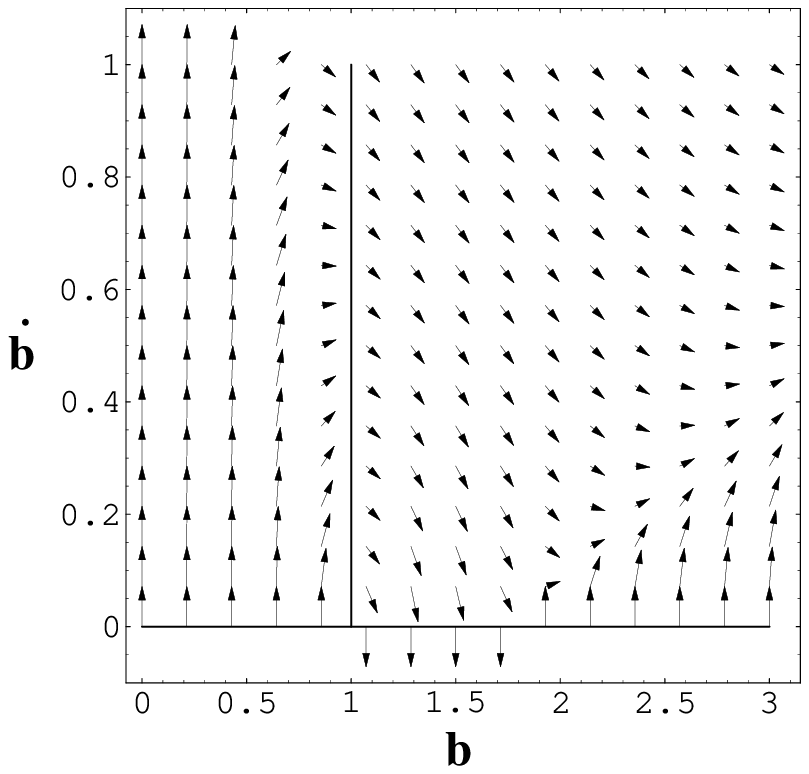}~~~
\includegraphics[height=5cm]
{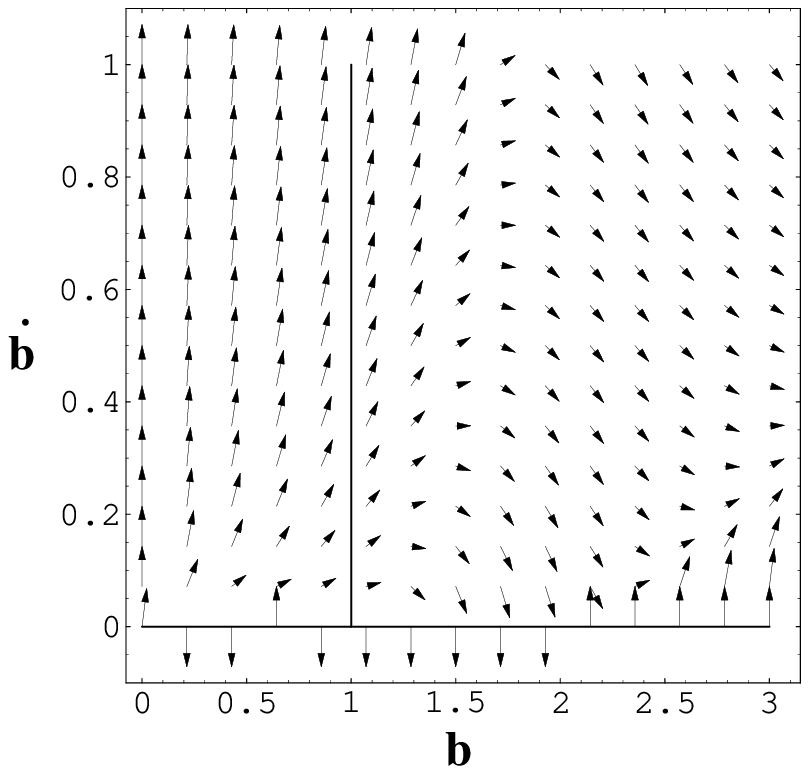}~~~
\includegraphics[height=5cm]
{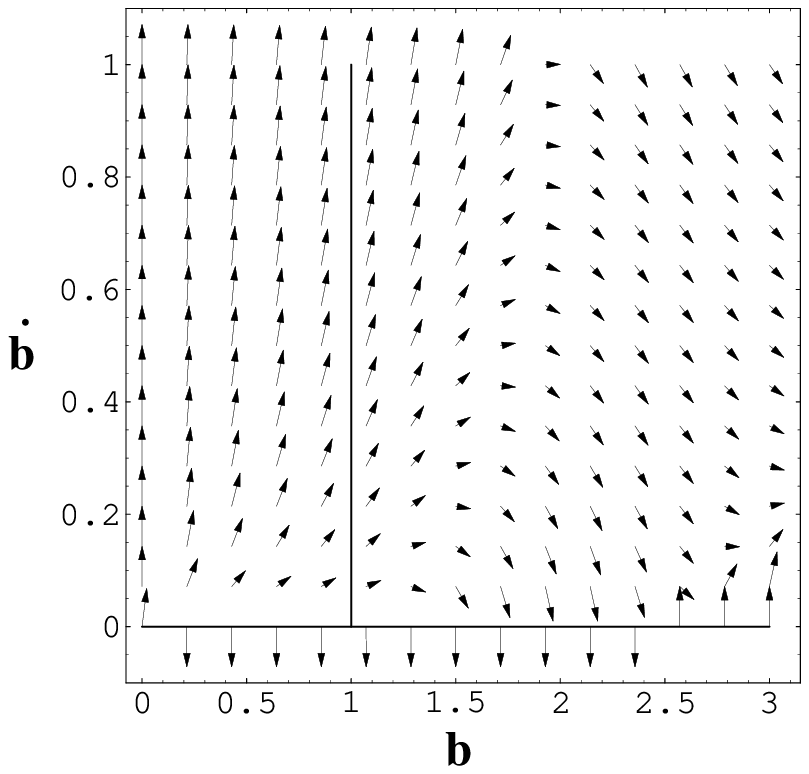}\\
~\qquad(a)\hspace{5cm}(b)\hspace{5cm}(c)
\caption{%
Phase space for model~2. We here take a unit such that $\Lambda=1$.   We
show (a) for $\gamma/\alpha=0$, (b) for $\gamma/\alpha=1$, (c) for
$\gamma/\alpha=2$, respectively.}
\label{fig1}
\end{figure}

\section{Summary and prospects}
\label{seclast}
In this paper, we have considered special cases for extension of
Einstein--Maxwell theory with higher order terms.
The models proposed in the present paper have the same solution
for spontaneous compactification $M_{D-2}\otimes S^2$ utilizing the
internal flux as the RSS model. The stability of the compactified vacuum
have been confirmed at the classical level. 
Nevertheless, the models reveal the different cosmological behaviors
because of the additional degrees of freedom or the existence of
the additional de Sitter phases.
We should investigate evolutional equations by using numerical
calculations in future work.

Our toy models seem to be quite special, but the analysis of the models
will exhibit some typical behaviors including inflation, and we can regard
general higher order corrections as a modification from the critical
models.

Recently, the black holes in higher derivative gravity attract
renewed interest \cite{LPPS,KKLR}. We consider that our model~1 is
appropriate to examine magnetized black holes of a novel type.

The subjects we should consider are, for instance: the coupling to matter
fields; compactifications on $(S^2)^N$ \cite{BDM} in our models;
the inclusion of a $p$-form flux field such as the Freund--Rubin
compactification \cite{FR}; the treatment of our models in quantum
cosmology
\cite{Halliwell,ZL1,ZL2}; the initial singularity problems
\cite{Starobinsky} in the models; the possible supersymmetrization of the
models; the field content of the model in
$3+2$ dimensions.

The generalization of our model~1 to more higher order theory is
considered straightforwardly. If we introduce an auxiliary field
$S_{M}^{N}$, the additional term can be replaced by
\begin{equation}
n\delta^{M_1\cdots M_{n-1}M}_{N_1\cdots N_{n-1}N}
S_{M_1}^{N_1}\cdots
S_{M_{n-1}}^{N_{n-1}}(\Delta_M^N{}^{PQ}T_{PQ}-S_M^N)
+\delta^{M_1\cdots M_n}_{N_1\cdots N_n}S_{M_1}^{N_1}\cdots
S_{M_n}^{N_n}\,,
\end{equation}
which will be studied extensively elsewhere.

Finally, we must research the relation to other theories.
We suppose that the structure of the model~1 owes to the matrix
$\Delta_{MNPQ}$, which appears also in the spin-two propagator. 
It would be very interesting if our models were the effective theory of
another theory.




\bibliographystyle{apsrev4-1}




\end{document}